\documentclass[12pt]{iopart}

\usepackage{amsfonts,amssymb}
\usepackage{graphicx,xcolor}
\usepackage{upgreek}
\usepackage{ulem}
\usepackage[utf8]{inputenc}
\usepackage{textcomp}

\usepackage[sorting=none]{biblatex}
\begin{document}

\title[On-Shot Plasma Mirror Characterization]{On-Shot Characterization of Single Plasma Mirror Temporal Contrast Improvement}

\author{L. Obst$^{1,2}$, J. Metzkes-Ng$^{1}$, S. Bock$^{1}$, G. E. Cochran$^3$, T. E. Cowan$^{1,2}$, T. Oksenhendler$^4$, P. L. Poole$^5$, I. Prencipe$^{1}$, M. Rehwald$^{1,2}$, C. R\"odel$^6$, H.-P. Schlenvoigt$^{1}$, U. Schramm$^{1,2}$, D. W. Schumacher$^3$, T. Ziegler$^{1,2}$ and K. Zeil$^{1}$}

\address{$^{1}$Helmholtz-Zentrum Dresden\,-\,Rossendorf (HZDR), Bautzner Landstr. 400, 01328 Dresden, Germany}
\address{$^{2}$Technische Universit\"at Dresden, 01062 Dresden, Germany}
\address{$^{3}$The Ohio State University, 191 West Woodruff Ave, Columbus, OH 43210, USA}
\address{$^{4}$iTEOX, 14 Avenue Jean Jaurès, 91940 Gometz-le-Châtel, France}
\address{$^{5}$Lawrence Livermore National Laboratory, 7000 East Ave, Livermore, CA 94550, USA}
\address{$^{6}$Friedrich-Schiller-Universit\"at Jena, Max-Wien-Platz 1, 07743, Jena, Germany}

\ead{l.obst@hzdr.de}
\begin{abstract}
We report on the setup and commissioning of a compact recollimating single plasma mirror for temporal contrast enhancement at the Draco 150 TW laser during laser-proton acceleration experiments.
The temporal contrast with and without plasma mirror is characterized single-shot by means of self-referenced spectral interferometry with extended time excursion (SRSI-ETE) at unprecedented dynamic and temporal range.
This allows for the first single-shot measurement of the plasma mirror trigger point, which is interesting for the quantitative investigation of the complex pre-plasma formation process at the surface of the target used for proton acceleration.
As a demonstration of high contrast laser plasma interaction we present proton acceleration results with ultra-thin liquid crystal targets of $\sim 1\ \upmu$m down to 10 nm thickness.
Focus scans of different target thicknesses show that highest proton energies are reached for the thinnest targets at best focus.
This indicates that the contrast enhancement is effective such that the acceleration process is not limited by target pre-expansion induced by laser light preceding the main laser pulse.

\end{abstract}

\maketitle

\section{Introduction}

Laser-proton acceleration is critically dependent on laser temporal contrast conditions \cite{Kaluza2004, Ceccotti07,Zeil2012, Andreev2009}.
Numerous experimental studies find that improving the temporal contrast before the arrival of the main laser pulse inhibits significant target pre-expansion which allows for the use of ultra-thin targets, that, until a certain lower limit (often referred to as "optimal thickness"), deliver increased proton energies \cite{Steinke10,Poole2018,Bin2017}.
However, quantitatively linking proton cut-off energies to the laser temporal contrast is the subject of ongoing highly complex experimental efforts in the field.\newline
Plasma mirror (PM) setups in the laser beam line are widely established tools to improve the laser temporal contrast by few orders of magnitude \cite{Ziener2003,Doumy2004,Roedel2011,Dromey2004} with the tradeoff of losing up to 50 \% of laser energy in the process. 
Implementation, relying on typical operation intensities exceeding 10$^{16}$ W/cm$^2$, can either be realized in the final focusing beam line directly before laser-target interaction \cite{Doumy2004APB} or in a separate setup, where two focusing optics are necessary to focus and recollimate the laser beam before and after interaction with the PM substrate, respectively.
As the laser impinges on the PM surface it is mainly transmitted as long as the intensity is below the ionization threshold of the substrate material.
Once the intensity rises above the ionization threshold, the surface becomes a plasma which reflects the laser beam. 
Under optimal conditions this trigger point lies few 100 fs before the peak of the laser pulse resulting in a suppression of amplified spontaneous emission (ASE) and pre-pulses by few orders of magnitude and an overall steepening of the rising slope of the main laser pulse.
The maximum fluence on the PM substrate can be varied by its translation along the focused laser beam which in turn determines the time of the trigger point before the main laser pulse.
\newline
We present the combined implementation of a single plasma mirror and the on-shot characterization of its temporal contrast enhancement performance, at unprecedented dynamic and temporal range, in the laser-proton acceleration experiment at the Draco 150 TW laser. 
This is enabled by analyzing a small part of the high-power pulse by means of self-referenced spectral interferometry at extended time excursion \cite{Oksenhendler2017,Oksenh2001} (SRSI-ETE), while the major portion of the laser beam is used for the laser-proton acceleration experiment.
Simultaneous measurement of the rising slope of the main laser peak over eight orders of magnitude and the resulting laser driven proton beam parameters is therefore possible for the first time.
The temporal contrast characterization is complemented on a broader temporal range by a scanning third order auto-correlator (Sequoia by Amplitude Technologies), allowing for multi-shot contrast measurements on the few 100 ps-timescale.
In combination with the available probing techniques (not subject of this work but details can be found in \cite{Metzkes2016,Ziegler2018}), this setup sets the stage for in-depth investigation of the particular target density distribution that results from interaction with laser light preceding the main laser peak (i. e. ASE, pre-pulses and the rising slope on the ps-timescale), and the resulting proton beam characteristics after interaction with the main pulse.
\newline
This work focuses on the implementation and commissioning of a recollimating single plasma mirror and its temporal contrast enhancement characterization with the SRSI-ETE. 
We find that ASE and pre-pulses are suppressed by 2-3 orders of magnitude before initiation of the plasma mirror triggering at $\sim -1$ ps, and we are able to deduce the final PM switching with high precision on a single-shot basis at $\sim -200$ fs before the arrival of the main laser pulse.
A fluence scan of the PM substrate along the laser axis is presented and compared to the expansion model given in \cite{Shaw2016} to estimate the expansion velocity, ionization intensity and absorption at the PM surface. 
Finally, we demonstrate that the achieved laser contrast improvement allowed for the use of ultra-thin targets of few ten nm thickness for laser-proton acceleration, resulting in increased cut-off energies as compared to the micrometer-thick reference targets \cite{Steinke10,Poole2018}.

\section{Experimental Setup}

\begin{figure}
\centering
\includegraphics[width=0.7\textwidth]{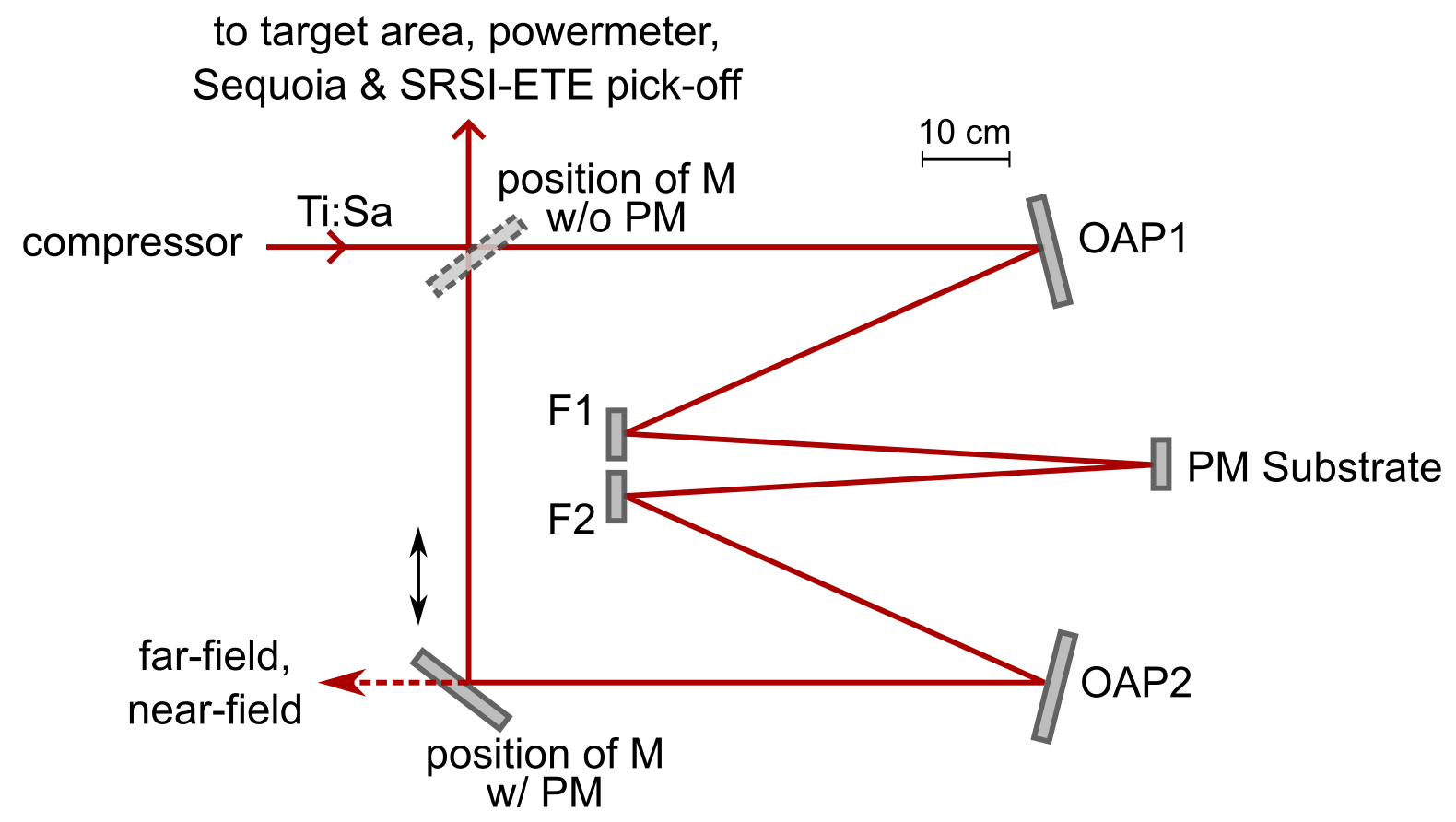}
\caption{Plasma mirror setup: after compression the laser pulse enters the experimental chamber where it can either be sent to the target area directly or first propagate through the plasma mirror beam line, depending on the position of the motorized mirror M. Folding the beam between the off-axis parabolas OAP1 and OAP2 and the plasma mirror substrate via F1 and F2 allows for a compact setup with a footprint no larger than 1 m$^2$ at a (collimated) beam diameter of 10 cm.}
\label{fig:setup}
\end{figure}

The high-power Titanium:Sapphire based laser system Draco (Dresden laser acceleration source) delivers ultrashort pulses with 30 fs pulse length and a peak power of 150 TW at a maximum repetition rate of 1 Hz \cite{Schramm2017}. 
After final pulse compression and wavefront correction by means of a deformable mirror, the p-polarized laser pulse enters the experimental chamber dedicated to laser-proton acceleration (see Fig. \ref{fig:setup}). 
There, a turning mirror (M) can be positioned on demand to allow for direct laser propagation to the target area ("M w/o PM") or for temporal contrast enhancement by propagation through the recollimating PM setup ("M w/ PM"). 
In the latter case, the laser pulse is focused by an F/10 off-axis parabolic mirror (OAP1), folded by the mirror F1, reflected off the PM substrate, folded by the mirror F2 and recollimated by the F/10 OAP2. 
F1 and F2, while allowing for a compact setup (by saving approximately 20 cm along the focusing direction, i.e. the horizontal dimension in Fig. \ref{fig:setup}), need to withstand laser fluences of 120 mJ/cm$^2$.
In the given setup no debris shielding of the OAPs is necessary as they face away from the plasma plume ejected after laser-PM interaction.
As PM substrates serve rectangular glass substrates of dimensions 100 $\times$ 50 mm$^2$ with an anti-reflection coating according to the spectral bandwidth of the laser pulse of $\sim$ 760-840 nm. 
Alignment at low energies is performed via a narrow high-reflection coated column on one side of the substrate.
Motorized linear stages allow for the translation of the PM substrate to a fresh position after each laser shot.
Alignment of the PM substrate mount comprised verification that the final focus position at the target for laser proton acceleration was not affected by translation of the PM substrate.
\newline
On-shot monitoring of the intensity mode of the collimated reflected beam and its pointing stability is realized by focusing the leakage through mirror M with an F/10 singlet lens into a diagnostic beam line with CCD cameras recording the laser near- and far-field,\footnote{In the given setup the on-shot far-field could not serve to monitor final focusing quality, as the wave front correction was performed for the high intensity focus at the target, while the on-shot far-field was at a different position along the beam line, for which the wave front was not optimized to the same quality level.} respectively.
Before entering the target area for proton acceleration, a full-aperture powermeter can be inserted into the laser beam to measure the laser pulse energy in vacuum.
The temporal contrast measurement is performed with a fraction of the laser beam that is picked off with a half-inch mirror positioned just within the outer boundary of the collimated beam profile after mirror M.
This allows for applying the major part of the laser pulse for proton acceleration.
The diagnostic beam is transported through a 1 mm thin window and distributed either to the SRSI-ETE or the Sequoia by a flipping mirror.
While the Sequoia requires at least $\sim$ 100 shots to scan the accessible temporal window with sufficient statistics, SRSI-ETE relies on tilted beam self-referenced spectral interferometry \cite{Oksenhendler2017}, enabling single-shot characterization of the pulse contrast with unprecedented intensity dynamic, temporal resolution and range.
Not only is the dynamic range largely sufficient to measure the contrast improvement induced by the plasma mirror of a few orders of magnitude, the increased temporal resolution also allows for close inspection of the main pulse's rising slope including the trigger point of the plasma mirror. 
\begin{figure}
\centering
\includegraphics[width=1.\textwidth]{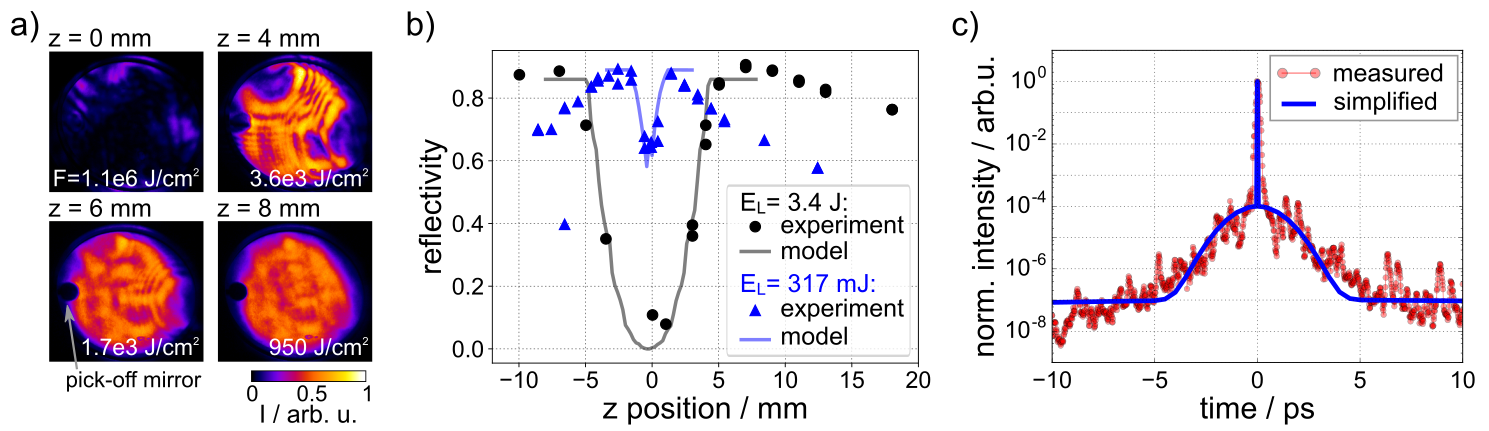}
\caption{Plasma mirror fluence scan. a) Laser near-field measurements after reflecting off the PM for different PM substrate positions along $z$ and the according fluences $F$. A small pick-off mirror was placed at the edge of the beam for diagnostic purposes that are not subject of this paper. b) PM reflectivity for entire fluence scan, as measured for two different laser energies: $E_L =$ 3.4 J (black markers) and $E_L =$ 317 mJ (blue markers). Lines show  plasma mirror expansion model from \cite{Shaw2016}. Comparison to our experimental results gives estimates on plasma parameters: $I_{ion}=6\times10^{12}$ W/cm$^2$, for $E_L$ = 3.4 J (317 mJ): ion sound speed $c_s=0.1\ \upmu$m/ps ($0.03\ \upmu$m/ps) and absorption of laser energy into plasma $\alpha=14$\% ($11$\%). c) Model input of simplified intrinsic laser temporal contrast curve (solid blue line) compared to Sequoia trace as measured before the PM (red markers).}
\label{fig:fluencescan}
\end{figure}
Powermeter, Sequoia and SRSI-ETE can be operated with and without passing the laser pulse through the PM setup, allowing for direct comparison between both configurations at full laser energy.
To compare the laser proton acceleration performance with and without temporal contrast enhancement, target focus scans were performed with liquid crystal film targets.
A linear slide target inserter (LSTI) \cite{Poole2016} forms films by drawing a small volume of the liquid crystal 8CB over a few mm diameter aperture.
Changes in wiper speed, frame temperature and liquid crystal volume result in film thicknesses ranging from few $\upmu$m down to 10 nm.
Films are formed at the same position ($\pm$ 2 $\upmu$m) well within the Rayleigh range of the laser focus ($\sim$ 30 $\upmu$m) and with similar surface conditions for every shot.
Final beam focusing onto the liquid crystal target was performed with an F/2.5 OAP under an incidence angle of 45\textdegree{} to a focal spot size of 3 $\upmu$m full width half maximum (FWHM), reaching laser intensities of $\sim$ 6 x 10$^{20}$ W/cm$^2$ (at a laser energy of 2.65 J on target).
Motorization of the target ensemble enabled laser intensity scans by translating the LSTI along the laser propagation direction.
Radio-chromic film stacks (RCF) and a Thomson Parabola spectrometer (TP) equipped with a multi-channel plate, both aligned in target normal direction, served as proton beam diagnostics \cite{Zeil2012}.

\section{Results}
In order to explore optimal operation conditions of the PM, a fluence scan was performed by translating the PM substrate several mm along the laser axis ($z$) around the laser focus of OAP1 and recording the reflected energy with the powermeter, while monitoring the reflected intensity mode with the near-field diagnostic behind mirror M. 
The results are displayed in Figure \ref{fig:fluencescan} a) and b).
In agreement with the literature on plasma mirrors, the reflectance of the PM is at a minimum when positioned in the laser focus at $z = 0$, due to strong perturbation of the plasma surface at the given intensity of up to 10$^{20}$ W/cm$^2$ (spot size $\sim$ 20 $\upmu$m FWHM, Rayleigh range $\sim$ 250 $\upmu$m).
Performing the fluence scan at a lower laser energy of 317 mJ results in a narrower reflectivity drop and allows for a more precise measurement of the best focus position.
At full energy, defocusing by $z\ \pm\ 7$ mm leads to reflectivities reaching a maximum of 90 \% and decreasing for larger distances from the focus.
The displayed images of the reflected laser near-field suggest an unperturbed laser intensity mode, and hence optimal final focusability on target, for distances of $z \geq 8 $ mm
This corresponds to spot diameters $\geq 680\ \upmu$m and intensities $\leq 3 \times 10^{16}$ W/cm$^2$ (i.e. fluences $\leq 950$ J/cm$^2$) on the PM substrate.
We compare the results of our fluence scan with the plasma mirror model from \cite{Shaw2016,Scott2015}, which allows for the derivation of important plasma parameters from the specific shape of the reflectivity drop around $z = 0$ mm (ref. to Fig. \ref{fig:fluencescan} b).
The model calculates the PM reflectivity based on the laser energy absorbed in the plasma, the shape of the plasma surface when the main laser peak is reflected and the intensity distribution of the laser spot on the PM.
While the near-field images of the reflected beam imply partial expansion and small-scale perturbation of the PM surface (e.g. $z$ = 4 mm in Fig. \ref{fig:fluencescan} a) \cite{Metzkes2016}, the PM model relies on a more homogeneous, normally distributed expansion. 
Using a simplified version of the laser temporal contrast curve as an input for the calculation (Fig. \ref{fig:fluencescan} c), we derive estimates for the ionization threshold intensity of the substrate atoms of  $I_{\mathtt{ion}}\sim 6\times10^{12}$ W/cm$^2$, expansion velocity of the plasma mirror surface $c_s \sim 0.1\ \upmu$m/ps and the absorbed laser energy fraction into the plasma $\alpha \sim 14$\% for an input laser energy of 3.4 J.
The derived parameters are in good agreement with the literature on plasma mirrors and femtosecond laser ionization of solids \cite{Dromey2004, Mora2003, Scott2015, Vu1994}.
\newline
For the results presented in the following, $z = 8$ mm was selected as working point, for which we show the plasma mirror performance at minimized laser energy loss while ensuring an unperturbed plasma surface when the main pulse is reflected, which was monitored on every shot in the near-field diagnostic. 
At this value of $z$, a mean laser intensity of 2-3 $\times 10^{16}$ W/cm$^2$ is estimated on the PM substrate resulting in the observed reflectivity of 89 \%. 
\begin{figure}
\centering
\includegraphics[width=\textwidth]{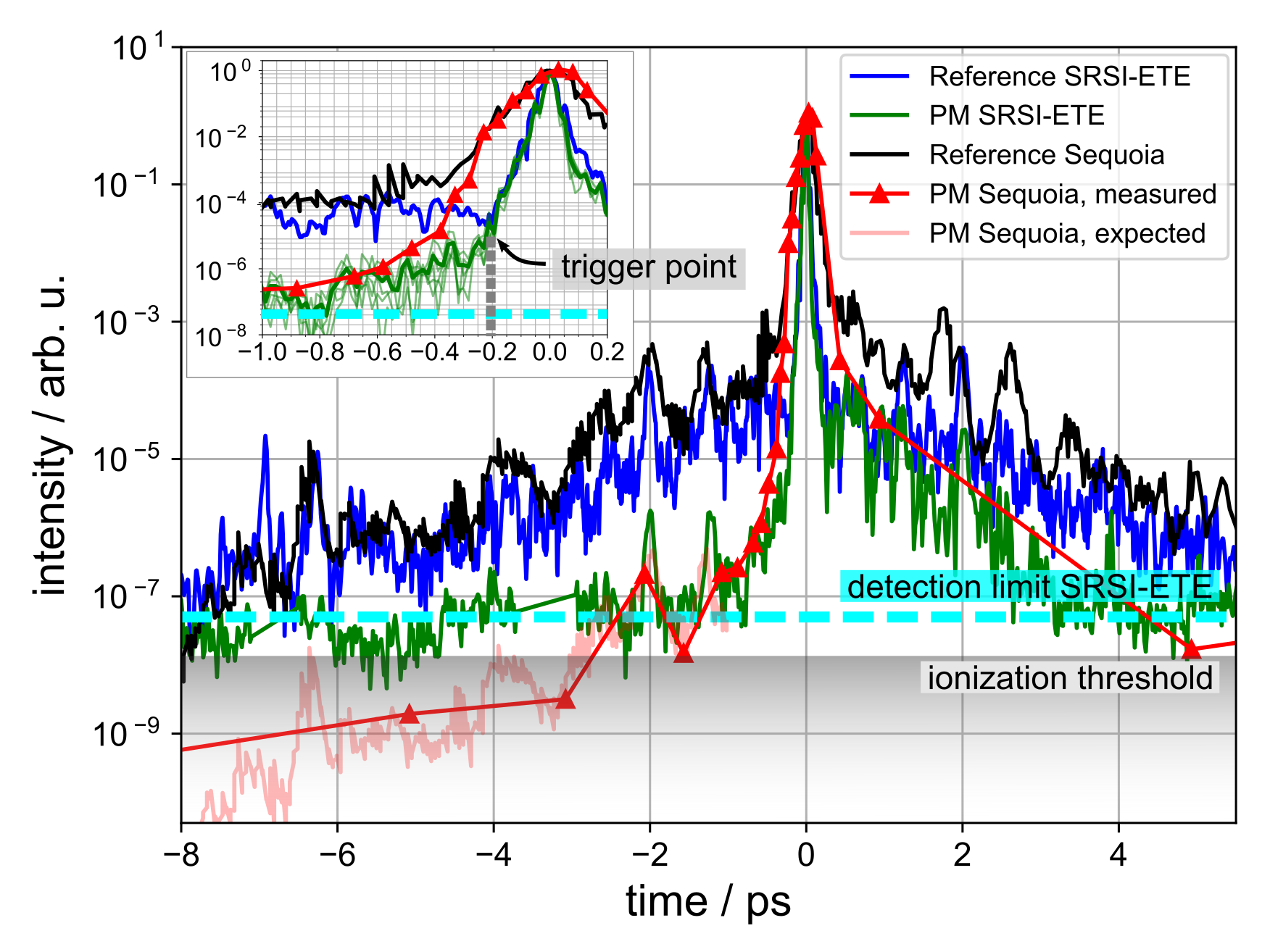}
\caption{Laser temporal contrast traces comparing the intrinsic laser contrast, measured with the Sequoia (black) and the SRSI-ETE (blue) with the PM-enhanced contrast (red and green, respectively). The thin red line indicates the complete course of the enhanced contrast curve at times $\leq -1$ ps, and is the result of scaling the reference curve with the contrast enhancement factor, which is deduced from the Sequoia measured data points with PM (which are the result of averaging over 3-4 shots each). The SRSI-ETE traces are averaged over 5 shots. Both measurement techniques agree well within the dynamic range of the SRSI-ETE ($5 \times 10^8$ in the given setup, indicated by light blue dashed line), while the SRSI-ETE offers a higher temporal resolution, allowing for close inspection of the main pulse's rising slope and the PM trigger point on a single-shot basis (view inset, where the averaged trace is overlaid with a number of consecutive single-shot measurements, each represented by a thin green line). On the other hand, the Sequoia trace gives the crucial result that the PM inhibits target ionization (grey shaded area) up to $~\sim$ 3 ps before the arrival of the main laser pulse. The entire temporal range of the Sequoia is not displayed here for clarity but can be found in \cite{Schramm2017}.}
\label{fig:contrast}
\end{figure}
Figure \ref{fig:contrast} displays the achieved temporal contrast improvement measured on a single-shot basis with SRSI-ETE and on a multi-shot basis with the Sequoia.
While the dynamic range of SRSI-ETE of $5 \times 10^8$ in the given setup is roughly 3-4 orders of magnitude lower than of the Sequoia, its temporal resolution is significantly better ({$\sim$ 20 fs as compared to 100 fs}), thus enabling higher precision measurements of the rising slope of the main laser pulse.
Within the intensity dynamic range of the Sequoia ($\sim$ 10$^{10}$), the intrinsic laser contrast (before contrast cleaning with the PM) is better or equal to $10^{-10}$ at $-120$ ps \cite{Schramm2017} (not displayed here for clarity) before the arrival of the main laser pulse.
Pre-pulses that are present on the $-100$ to $-10$ ps-timescale before arrival of the main laser pulse are efficiently suppressed by the plasma mirror and their remaining intensity after the PM is below the ionization threshold of the target for laser proton acceleration.
The contrast improvement achieved with the PM can be deduced at the rising flank of the main laser pulse: at $-1.6$ ps both Sequoia and SRSI-ETE measurements show that the relative laser intensity is lowered from $2 \times 10^{-5}$ to $3 \times 10^{-8}$, suggesting an improvement of almost 3 orders of magnitude.\newline
The inset in Fig. \ref{fig:contrast} shows the contrast development close to the trigger point: while the distance between the curves with and without PM already starts to decrease at $\sim - 1$ ps, indicating the initiation of the PM triggering, the curves merge at $-200$ fs which is the actual switching point after which the PM causes no further contrast improvement.\footnote{The laser intensity at the position of the pick-off mirror for contrast measurement is slightly lower (40\% as deduced from the recorded near field images) than in the center of the beam.  This results in a systematic measurement error of the PM triggering point of only few fs as deduced from the PM model \cite{Shaw2016}, which, compared to the temporal resolution of SRSI-ETE of 20 fs, can be omitted.}
Of both devices, the SRSI-ETE gives the most accurate measurement of the plasma mirror effect on the rising slope of the laser, showing remarkable shot-to-shot stability of the PM switching (thin green lines in the inset of Fig. \ref{fig:contrast}), while the Sequoia measurement is clearly limited by its comparably low temporal resolution.
On the other hand, from the Sequoia trace follows that when irradiating the target with a PM-cleaned pulse, ionization of the target surface starts no earlier than $\sim\ 3$ ps prior to the arrival of the main laser pulse.\newline
Future PM implementations at the Draco laser may comprise two plasma mirror substrates \cite{Levy2007} to further improve the laser contrast.
The effect of debris produced at one of the substrates and deposited on the second and its impact on the contrast enhancement performance would certainly have to be investigated over the course of several shots, which is a straightforward measurement with SRSI-ETE.\newline
\begin{figure}
\centering
\includegraphics[width=0.7\textwidth]{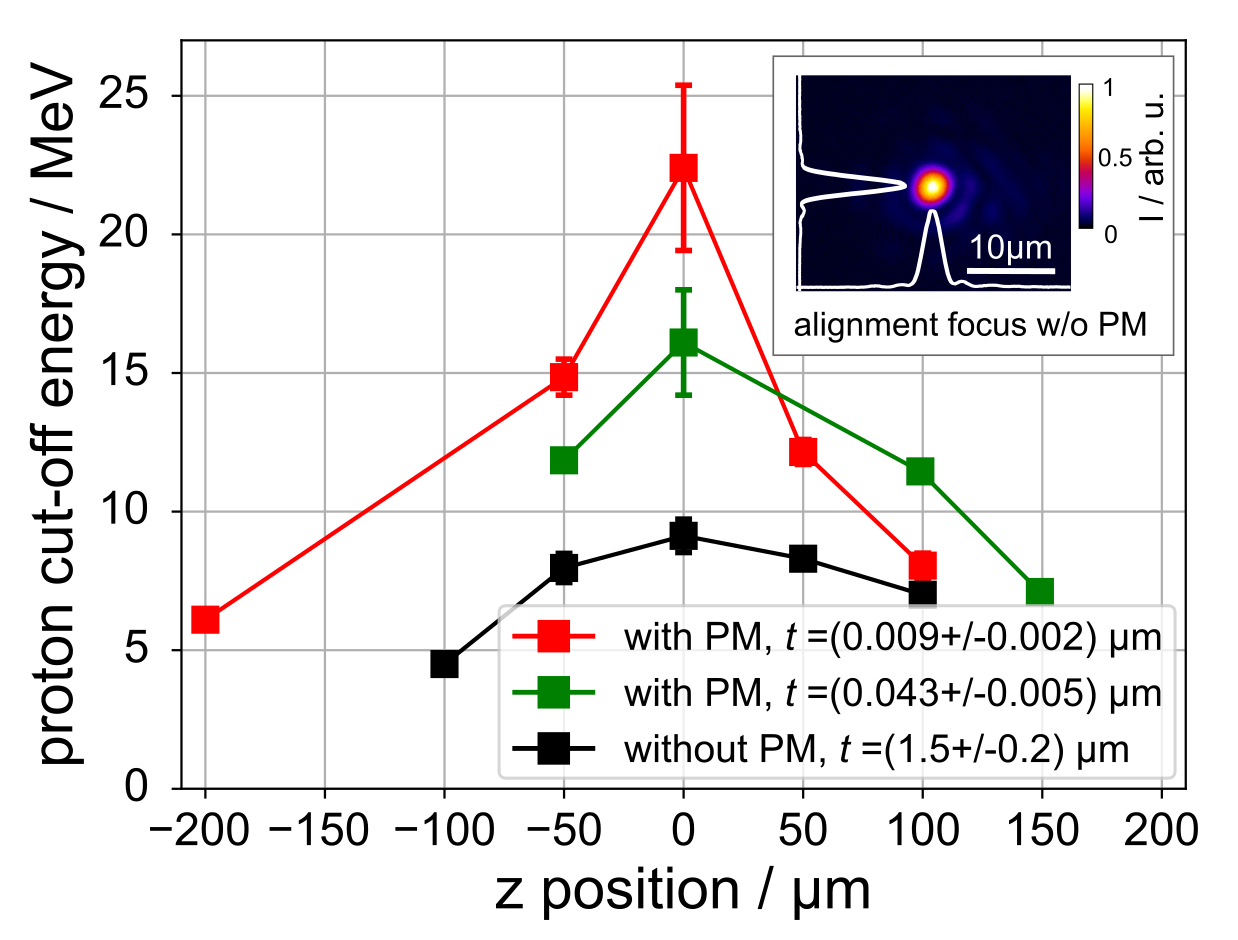}
\caption{Target focus scan with and without PM. Shots with PM on target thicknesses $t$ between 7 and 11 nm and 38 and 48 nm are averaged. Without plasma mirror shots on target thicknesses between 1.3 and 1.7 $\upmu$m. The inset shows the optimized low energy focus without PM after wave front correction.}
\label{fig:focus}
\end{figure}

Figure \ref{fig:focus} displays the proton acceleration results with ultra-thin liquid crystal film targets under oblique incidence (details of the experiment and the full target thickness scan can be found in \cite{Poole2018}) in the form of a target focus scan.
The target position was monitored before every shot with a scattered light imaging system that was referenced to an alignment target imaged and placed in focus via the focus diagnostic.
The scattered light imaging system confirmed that indeed every film was formed in the same plane with respect to the best focus, thus ensuring comparability between shots on different target thicknesses ($t$) with and without PM. 
We find that proton energies are increased at least two-fold when going from $t\sim 1\ \upmu$m at intrinsic laser contrast to $t \sim 10$ nm with PM-enhanced contrast.
The observed tendency of increasing proton cut-off energies with decreasing target thickness is well understood:
hot electrons originating from the focus region travel ballistically through the target with a certain beam divergence before they exit the target rear surface to form an electron sheath.
Thinner targets result in a smaller diameter of this sheath and hence, an increased accelerating field acting on the protons and ions in the rear-surface contamination layer (ref. to \cite{Daido2012} and references therein).
However, the minimum target thickness is usually limited by the experimental laser contrast (e.g. \cite{Bin2017}).
In our case, without PM no proton signal was detected for targets below 300 nm, indicating significant target pre-expansion and subsequent decay of acceleration gradients.\footnote{It should be noted that the intrinsic temporal contrast of the double CPA upgrade of Draco 150 TW has been significantly improved since the presented study was conducted. Since then, proton energies $\sim$ 20 MeV have again been demonstrated without plasma mirror from 2 $\upmu$m thick Ti foils \cite{Obst2017}, similar to earlier results with an optimized single CPA system \cite{Metzkes2014}.}\newline
While both cases - PM contrast with thin targets and intrinsic contrast with thick targets - show highest proton energies at the highest intensity focus position ($z=0$ in Fig. \ref{fig:focus}), a steeper trend is observed for the PM contrast case, specifically for the thinnest targets used, when approaching $z = 0$.
From the target focus scan we can further conclude that no significant focusing or defocusing term is added when the laser reflects off the plasma mirror surface, which would result in a systematic offset of the best focus position as compared to shots at intrinsic laser contrast.
This indicates a flat, un-expanded plasma mirror surface, consistent with the recorded near-field images and suggesting optimal focusability which in turn is in agreement with the observed high proton energies. 

\section{Conclusion}
We presented the implementation and characterization of a compact recollimating single plasma mirror setup that can be used on demand for laser-proton acceleration experiments at the Draco 150 TW laser.
Single-shot characterization of the temporal contrast is implemented with the SRSI-ETE \cite{Oksenhendler2017}, allowing for the inspection of the rising slope of the laser pulse at unprecedented dynamic, temporal range and resolution.
An important result is the determination of the plasma mirror switching point on a single-shot basis at $(-200\pm 20)$ fs before the main laser pulse, which to our knowledge is the first measurement of this kind \cite{Itakura2015}.
In a proton acceleration experiment with the contrast-cleaned Draco 150 TW, an increase in proton energies by roughly a factor two was observed when going from $\sim 1\ \upmu$m thick to $\sim 10$ nm thin liquid crystal film targets.\newline
The presented single-shot contrast improvement and characterization setup will enable developing a deeper understanding, i.e. by generating input parameters for numerical studies, of the influence of the laser pulse's rising slope on the ps-timescale on the target surface pre-plasma density scale length, and the resulting effect on the laser proton acceleration performance.
Extending these studies to even higher laser intensities is possible due to the recent implementation of a similar plasma mirror setup layout including SRSI-ETE diagnostic in the Petawatt branch of the Draco laser \cite{Schramm2017}.


\section*{Acknowledgment}
R. Gebhardt and U. Helbig are highly acknowledged for their laser support.
The work was partially supported by EC Horizon 2020 LASERLAB-EUROPE / LEPP (Contract No. 654148) and by the NNSA under Contract DE-NA0003107. 
C. R\"odel acknowledges financial support from the Volkswagen Foundation. 

\section*{Licence}
This Accepted Manuscript is available for reuse under a CC BY-NC-ND 3.0 licence after the 12 month embargo period provided that all the terms of the licence are adhered to.

\printbibliography


\end{document}